\documentclass[useAMS,usenatbib]{mn2e}
\usepackage{url,times,graphicx,amsmath,amsfonts,amssymb,aas_macros,color,epsfig,epstopdf}
\usepackage{xcolor}

\def \etal {et~al.~}

\newcommand{\hMpc}{{\ifmmode{h^{-1}{\rm Mpc}}\else{$h^{-1}$Mpc}\fi}}
\newcommand{\hkpc}{{\ifmmode{h^{-1}{\rm kpc}}\else{$h^{-1}$kpc}\fi}}
\newcommand{\kpc}{{\ifmmode{ {\rm kpc} }\else{{\rm kpc}}\fi}}
\newcommand{\kms}{{\ifmmode{ {\rm km\,s^{-1}} }\else{ ${\rm km\,s^{-1}}$ }\fi}}
\newcommand{\hMsun}{{\ifmmode{h^{-1}{\rm {M_{\odot}}}}\else{$h^{-1}{\rm{M_{\odot}}}$}\fi}}
\newcommand{\Msun}{{\ifmmode{{\rm M}_{\odot}}\else{${\rm M}_{\odot}$}\fi}}
\newcommand{\Mhalo}{{\ifmmode{M_{\rm halo}}\else{$M_{\rm halo}$}\fi}}
\newcommand{\Rvir}{{\ifmmode{R_{\rm vir}}\else{$R_{\rm vir}$}\fi}}
\newcommand{\Mstar}{{\ifmmode{M_{\star}}\else{$M_{\star}$}\fi}}
\newcommand{\Vrot}{{\ifmmode{V_{\rm rot}}\else{$V_{\rm rot}$}\fi}}
\newcommand{\ltsima}{$\; \buildrel < \over \sim \;$}
\newcommand{\gtsima}{$\; \buildrel > \over \sim \;$}
\newcommand{\lsim}{\lower.5ex\hbox{\ltsima}}
\newcommand{\gsim}{\lower.5ex\hbox{\gtsima}}

\def\lesssim{\mathrel{\hbox{\rlap{\hbox{\lower4pt\hbox{$\sim$}}}\hbox{$<$}}}}
\def\gtrsim{\mathrel{\hbox{\rlap{\hbox{\lower4pt\hbox{$\sim$}}}\hbox{$>$}}}}

\newcommand{\beq}{\begin{equation}}
\newcommand{\eeq}{\end{equation}}
\def\beqa{\begin{eqnarray}}
\def\eeqa{\end{eqnarray}}
\def\LCDM{\ensuremath{\Lambda}CDM}

\def\head{ \vbox to 0pt{\vss \hbox to 0pt{\hskip 440pt\rm
      LA-UR-10-07069\hss} \vskip 25pt}}
\def \xoff {\ifmmode x_{\rm off} \else $x_{\rm off}$ \fi}
\def \rhorms {\ifmmode \rho_{\rm rms} \else $\rho_{\rm rms}$ \fi}

\def \kms {\ifmmode  \,\rm km\,s^{-1} \else $\,\rm km\,s^{-1}  $ \fi }
\def \kpc {\ifmmode  {\rm kpc}  \else ${\rm  kpc}$ \fi  }  
\def \hkpc {\ifmmode  {h^{-1}\rm kpc}  \else ${h^{-1}\rm kpc}$ \fi  }  
\def \hMpc {\ifmmode  {h^{-1}\rm Mpc}  \else ${h^{-1}\rm Mpc}$ \fi  }  
\def \Mpch {\ifmmode  {h^{-1}\rm Mpc}  \else ${h^{-1}\rm Mpc}$ \fi  }  
\def \Msun {\ifmmode {\rm M}_{\odot} \else ${\rm M}_{\odot}$ \fi} 
\def \hMsun {\ifmmode h^{-1}\,\rm M_{\odot} \else $h^{-1}\,\rm M_{\odot}$ \fi}

\def \LCDM {\ifmmode \Lambda{\rm CDM} \else $\Lambda{\rm CDM}$ \fi}
\def \sig8 {\ifmmode \sigma_8 \else $\sigma_8$ \fi} 
\def \OmegaM {\ifmmode \Omega_{\rm m} \else $\Omega_{\rm m}$ \fi} 
\def \Omegab {\ifmmode \Omega_{\rm b} \else $\Omega_{\rm b}$ \fi} 
\def \OmegaL {\ifmmode \Omega_{\rm \Lambda} \else $\Omega_{\rm \Lambda}$\fi} 
\def \Deltavir {\ifmmode \Delta_{\rm vir} \else $\Delta_{\rm vir}$ \fi}
\def \rhocrit {\ifmmode \rho_{\rm crit} \else $\rho_{\rm crit}$ \fi}
\def \rhou {\ifmmode \rho_{\rm u} \else $\rho_{\rm u}$ \fi}
\def \zc {\ifmmode z_{\rm c} \else $z_{\rm c}$ \fi}

\def\head{
.ps \vbox to 0pt{\vss
                   \hbox to 0pt{\hskip 440pt\rm LA-UR-10-07069\hss}
                  \vskip 25pt}}
\title[Strongly Coupled Cosmologies II] {Strongly Coupled
  Dark Energy Cosmologies: \\preserving ${\Lambda}$CDM success and easing low scale problems\\ II -
  Cosmological simulations}

\author[A.V. Macci\`o \etal]
       {Andrea V. Macci\`o$^1$\thanks{E-mail: maccio@mpia.de}, Roberto Mainini$^2$, Camilla Penzo$^1$, Silvio A.
         Bonometto$^3$\\
$^1$Max-Planck-Institut f\"ur Astronomie, K\"onigstuhl 17, 69117 Heidelberg, Germany\\
$^2$Milano--Bicocca University, Physics
  Department G.~Occhialini, Piazza della Scienza 3, 20126
  Milano-Bicocca\\
$^3$Trieste University, Physics Department, Astronomy Unit, Via
  Tiepolo 11, 34143 Trieste (Italy) \& I.N.A.F., Osservatorio
  Astronomico di Trieste}

\begin{document}

\date{Accepted XXXX . Received XXXX; in original form XXXX}

\pagerange{\pageref{firstpage}--\pageref{lastpage}} \pubyear{2010}

\maketitle

\label{firstpage}

%\clearpage

%%%%%%%%%%%%%%%%%%%%%%%%%%%%%%%%%%%%%%%%%%%%%%%%%%%
\begin{abstract}

  In  this  second  paper  we present  the  first  Nbody  cosmological
  simulations of strongly coupled Dark  Energy models (SCDEW), a class
  of models that alleviates theoretical issues related  to the nature
  of  dark energy. SCDEW models  assume a strong coupling between
  Dark Energy (DE)  and an ancillary Cold Dark  Matter (CDM) component
  together  with  the  presence  of  an  uncoupled  Warm  Dark  Matter
  component.   The strong  coupling between  CDM and  DE allows  us to
  preserve small scale fluctuations even if the warm particle is quite
  light ($\approx  100$ eV).  Our  large scale simulations  show that,
  for  $10^{11}<M/M_\odot<10^{14}$,  SCDEW  haloes  exhibit  a  number
  density  and distribution  similar to  a standard  Lambda Cold  Dark
  Matter   ($\Lambda$CDM)  model,   even   though   they  have   lower
  concentration parameters.  High resolution  simulation of a galactic
  halo ($M\sim 10^{12}  \Msun $) shows $\sim  60\%$ less substructures
  than its \LCDM counterpart, but  the same cuspy density profile.  On
  the scale of galactic satellites ($M\sim 10^{9} \Msun$) SCDEW haloes
  dramatically differ from  \LCDM. Due to the  high thermal velocities
  of the WDM component they are almost devoid of any substructures and
  present strongly  cored dark matter density  profiles. These density
  cores extend for several hundreds of parsecs, in very good agreement
  with Milky  Way satellites  observations.  Strongly  coupled models,
  thanks  to their  ability to  match observations  on both  large and
  small scales might  represent a valid alternative to  a simple \LCDM
  model.

\end{abstract}

%%%%%%%%%%%%%%%%%%%%%%%%%%%%%%%%%%%%%%%%%%%%%%%%%%%
\noindent
\begin{keywords}

cosmology: dark matter galaxies: evolution - formation methods:N-body
simulation

 \end{keywords}

%%%%%%%%%%%%%%%%%%%%%%%%%%%%%%%%%%%%%%%%%%%%%%%%%%%
\section{Introduction} \label{sec:introduction}
%%%%%%%%%%%%%%%%%%%%%%%%%%%%%%%%%%%%%%%%%%%%%%%%%%%

A large variety of data on both small and large scales points to the
existence of two dark components in our Universe: dark matter, which
dominates the gravitational budget of collapsed objects like galaxies
and clusters, and dark energy, which regulates the expansion of the
Universe at later times (Tegmark et al. 2006, Planck collaboration
2014). While there is a general agreement on the presence of these two
components, their nature is still very unclear and several
possibilities have been put forward in the last years. 

The most common model for Dark Matter (DM) and Dark Energy (DE) is the
Lambda Cold Dark Matter (\LCDM).  In this model DE behaves like a
cosmological constant, being often interpreted as vacuum energy, and
is therefore constant in space and time. Furthermore, DM particle
velocities are assumed to be negligible, as though they decoupled
quite early being already non--relativistic.

The \LCDM model is quite successful in reproducing data on large and
intermediate scales (e.g. Springel et al. 2005). On the other hand it
still raises several questions like the fine tuning problem, as the DE
energy density is $\sim (10^{-30}m_p)^4$ ($m_p:$ Planck mass), and the
coincidence or {\it ``Why now?''} problem: why DE, negligible during
all cosmic history, became significant just now and late enough to
allow non--linear structures to develop.

In the attempt to alleviate these problems, the option of DE being a
self--interacting scalar field with a tracker potential
was suggested (e.g., Ellis
et al, Wetterich 1988, Ratra \& Peebles 1988, Brax \& Martin 1999,
2000, see Ratra \& Peebles 2003 for a review).
Furthermore the possibility of a coupling between DM and DE
was deepened by several authors (e.g.
Wetterich 1995, Amendola 2000 Amendola \& Tocchini--Valentini 2001,
see Amendola \etal 2013 and references therein for a comprehensive review).

While tracker potentials are devoid of finely tuned scales,
the coupling allows for a non-negligible fraction of DE at early times
(e.g. Amendola 2000).
%(when, however, the very DM density tends to be negligible).
As expected, DM--DE coupling modifies the
equation of motion of a test particle in an expanding universe with
respect to the pure Newtonian case (Macci\`o et al. 2004).

In the past years several works studied the effect of a DM-DE coupling
on structure formation (e.g. Macci\`o et al. 2004, Caldera-Cabral et
al. 2009, Baldi et al. 2010, Li \& Barrow 2011, Baldi 2012).
After Planck data release, Xia (2014) outlined that the tension
between CMB and Hubble telescope $H_0$ estimates was eliminated by the
coupling option, finding a coupling constant $\beta =0.078\pm 0.022$. Other observables, however, are
not significantly modified by such coupling and this makes it hard to
disentangle coupled dark energy models from a more simple \LCDM one
(e.g. Baldi et al. 2012, Pace et al. 2015).

The \LCDM model also faces some tension with data on the scales of low
mass galaxies where the central cuspy dark matter distribution
predicted by CDM (e.g Navarro, Frenk \& White 1997) seems to be a very
poor match of the observed central cored dark matter distribution in
dwarf galaxies (e.g. Moore 1994, de Blok et al. 2001, Kuzio de Naray
et al. 2006, Oh et al. 2011, Salucci et al. 2012).

In the last years, there has been a mounting evidence that including the
effect of a dissipational baryonic component, strongly
modifies the central DM distribution (Governato et al. 2010, Macci\`o
et al. 2012a) reconciling observations and simulations on dwarf galaxy
scales (Di Cintio et al. 2014a).  

On the other hand these simulations have also shown that baryons are
effective in altering the DM density profile only up to a certain mass
scale (Governato et al. 2012, Di Cintio et al. 2014b, Arraki et
al. 2014, O{\~n}orbe et al. 2015) and that very low mass galaxies are
expected to retain a cuspy profile. 

Unfortunately this is in contrast with recent observations of the DM
distribution in the MW satellites, which seem to suggest the presence
of cored profiles even at these very low mass scales, where the effect
of baryons should be minimal. The presence of such cores might
indicate that DM is 
warm (e.g. Dalcanton \& Hogan 2001, Colin et al. 2008). Several works
have dealt with the effect of Warm Dark Matter (WDM) on halo structure
(Bode et al. 2001, Avila-Reese et al. 2001, Knebe et al. 2002,
Tikhonov et al. 2009, Schneider et al. 2012). Constraints on the mass
of a possible WDM candidate can be obtained from the analysis of the
matter power spectrum from the Lyman--$\alpha$ forest (Seljak et al. 2006, Viel
et al. 2005,2008, Boyarsky et al. 2009), current measurements suggest
the WDM mass (for a pure thermal candidate) to be around (or above)
3.5 keV (Viel et al. 2013, Polisensky \& Ricotti 2014).

A WDM particle of 3--4 keV, however, yields no improvement with
respect to CDM on small scales (Schneider et al. 2014). The situation
is particularly hopeless for halo density profiles, which require a
WDM particle of $\sim 100$ eV in order to create appreciable central
cores as the ones observed in the Milky Way satellites (Macci\`o et
al. 2012b, Shao et al. 2013).

In the light of these considerations it is  worth to explore
alternatives to \LCDM, aiming to a model which can deal, at the same
time, with both theoretical and observational issues.

In two recent papers (Bonometto et al. 2012, Bonometto \& Mainini
2014) a new model was indeed proposed, which combines DM-DE coupling
and Warm Dark Matter trying to overcome the problems of \LCDM, both on
the theoretical and observational side.  Observational DM, in this
model, is warm and light. An auxiliary CDM component, coupled to DE,
is however added, which never exceeds some permils of the total
density, but naturally exerts a number of key effects though cosmic
history.

In the companion paper (Bonometto, Mainini, Macci\`o 2015, Paper I
hereafter) we addressed the linear behavior of such models also
including a new option, that CDM--DE coupling fades at low $z$.  In
these models WDM is made of extremely low mass particles, down to few
tens of eV,
which might, in turn, have significant effects on the central DM
distribution in collapsed objects.

In this second paper, we aim to present the first Nbody simulations of
strongly coupled warm+cold models (SCDEW hereafter). Our goal is to
study these models in the highly non linear regime probed by structure
formation, both on large ($\sim $Mpc) and small ($\sim $100 pc)
scales.

The paper is organized as follows: we will first summarize the main
features of our novel cosmological model in section (\ref{sec:model}),
we will then introduce our numerical codes to generate initial
conditions and evolve the simulations (\ref{sec:simulation}), we will
then present results for large scales simulations and for zoomed
simulations of Milky Way-like objects and dwarf galaxies
({\ref{sec:Results}), we will then conclude with a discussion of our
  results and possible future developments (\ref{sec:conclusion}).

%%%%%%%%%%%%%%%%%%%%%%%%%%%%%%%%%%%%%%%%%%%%%%%%%%%
\section{Theoretical model} \label{sec:model}
%%%%%%%%%%%%%%%%%%%%%%%%%%%%%%%%%%%%%%%%%%%%%%%%%%%

In this Section we briefly review the main features of SCDEW
cosmologies, more extensively discussed in the previous associated
paper.
Besides of baryonic and radiative components ($\gamma$'s \& $\nu$'s),
these models assume
an ordinary uncoupled light WDM component, while
DE is coupled to an ancillary CDM component which, actually, never
plays the role that DM has in $\Lambda$CDM or similar models.

The starting point of SCDEW cosmologies is the finding that, during
radiative expansion, an attractor solution exists, for coupled CDM and
a scalar field components, keeping them a constant fraction of the
cosmic density in primeval radiative eras. In the absence of coupling,
these components would dilute $\propto a^{-3}$ and $\propto a^{-6}$
respectively. The time evolution of the different density parameters
as a function of the expansion factor $a$ in our SCDEW model is
presented in figure \ref{fig:omega}.

\begin{figure}
\includegraphics[width = 0.5\textwidth]{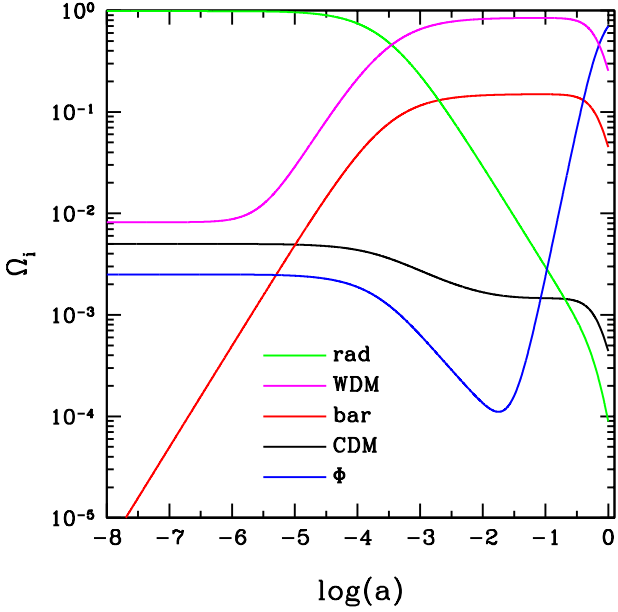}
\caption{Time evolution of the density parameters ($\Omega_i$)
  of radiation, Dark Energy ($\Phi$), Cold Dark Matter, Warm Dark Matter
  and baryons in the SCDEW model.}
\label{fig:omega}
\end{figure}

As shown in the Figure, the coupling, by allowing energy to flow from CDM
to the field, puts CDM and DE on an attractor solution where both shall dilute
$\propto a^{-4}$, as radiative components do. For a coupling constant
$\beta$, the fair constant primeval density parameters of the two
components, along the attractor, are $1/2\beta^2$ and $1/4\beta^2$,
respectively. Values $\beta \sim 10$ are favored, so that the early
contribution of CDM and field keeps steadily around or below $\approx 0.2\, \%$.
While causing no harm to BBN or CMB data, this contribution therefore keeps
non--negligible, possibly dating since the end of inflation.

When this stationary solution is broken by WDM derelativisation, the
Universe naturally evolves towards the observed features. In
particular, the field turns into the observed DE component.

Accordingly, at variance from $\Lambda$CDM models, in SCDEW models DE
has always been a small but non--negligible component, which eases fine
tuning 
while the coincidence problem is also attenuated, for a wide set
  of parameter choices, including those needed to obtain a reasonable
  fit to large scale data.  The option of the field $\Phi$ playing both the role of
inflaton and DE is also not excluded.

Coupled CDM however plays another key role. In the non--relativistic
regime, as already known (Amendola 2000), its coupling to DE causes a
strengthening of its self-gravity. This triggers a mechanism allowing
its fluctuations to restart WDM fluctuations, also on scales where
they had formerly been erased by free streaming, as soon as WDM
derelativizes. This revitalization of WDM fluctuations on scales below
the free streaming one allows us to have very ``hot'' WDM, of the order of
100 eV.

At more recent times ($z < \sim 50$), besides of keeping a density
parameter $< 10^{-1}$--$10^{-2}$ of baryons, CDM could dynamically
decouple from baryon and WDM fluctuations.

This decoupling might arise naturally, however in our current
parameterization we have imposed an {\it ad-hoc} decoupling at small
redshifts, which causes no substantial change in the expected
fluctuation spectra.
Of course a full consistent model should provide a mechanism
for the fading of the coupling at low redshift, which, for example,
can be a conseguence of the dynamics of the scalar field.
We refer the reader to Paper I (especially appendix A) for a thourough
discussion of this issue
while here we rather prefer to use a more phenomelogical approach.

For the sake of definiteness, therefore, the linear spectra used to
start the simulations in this work are those obtainable by setting a
decoupling parameter $d = 4$ (see Paper I), yielding a
full CDM--DE decoupling at $z \simeq 50$, where simulations are started.

%%%%%%%%%%%%%%%%%%%%%%%%%%%%%%%%%%%%%%%%%%%%%%%%%%%
\section{Numerical simulations} \label{sec:simulation}
%%%%%%%%%%%%%%%%%%%%%%%%%%%%%%%%%%%%%%%%%%%%%%%%%%%

Numerical simulations have been performed with the Nbody code {\sc
  pkdgrav} (Stadel \etal 2001), while the initial conditions have been
created with the {\sc grafic2} code (Bertchinger 2001), which we have
recently modified to allow a larger spectrum of cosmological models,
as described in Penzo \etal (2014).  The transfer function for the
power spectrum have been produced with a modified version of {\sc
  cmbfast} that allows for an explicit coupling between the cold dark
matter and the dark energy components (Bonometto \& Mainini 2014).

The redshift--zero power spectrum for our specific SCDEW model is
shown in figure \ref{fig:power}. Despite the presence of a very warm
component with a mass of 90 eV, the power spectrum does not differ
much from the expectation of a \LCDM model with the same cosmological
parameters. For the sake of comparison, in the same figure we also
show the power spectrum of a pure WDM model for the same warm particle
mass. As expected, it exhibit a dramatic reduction of the power on
small scales.

\begin{figure}
\includegraphics[width = 0.5\textwidth]{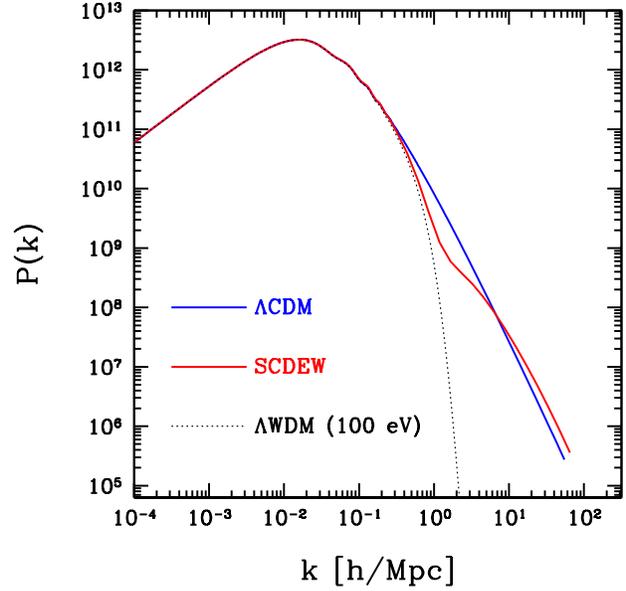}
\caption{Linear power spectrum at z=0 for the \LCDM model (blue) and
  the SCDEW model (red). The black thin line shows, for comparison,
  the power spectrum of a Lambda Warm Dark Matter model (LWDM) with a
  particle mass of 90 eV.}
\label{fig:power}
\end{figure}

As is evident from Figure \ref{fig:omega}, at the initial 
redshift of our simulations  ($z \approx 50$),
the CDM contribution to the gravitational potential is quite small.
When combined with our assumption of a vanishing coupling at late times
(see Paper I), this allows us to use the standard equations of motions
for the Warm Dark Matter components and to neglect the presence of a
Cold component in the simulation. On the other hand the initial
power spectrum and the evolution of the background do account
for the presence of all components, i.e., essentially,
the effects of a fair time dependence of Dark Energy.

Finally, in agreement with previous works on WDM (Colin \etal 2008,
Macci\`o \etal 2012b, Shao \etal 2013),
WDM particles were added the thermal velocity
\begin{equation}
\label{eq:velocities}
\frac{v_0(z)}{1+z}=.012\left(\frac{\Omega_{\rm w}}{0.3}\right)^{\frac{1}{3}}\left(\frac{h}{0.65}\right)^{\frac{2}{3}}\left(\frac{1.5}{g_{\rm w}}\right)^{\frac{1}{3}}\left(\frac{\text{keV}}{m_{\rm w}}\right)^{\frac{4}{3}}\kms
\end{equation}
in agreement with Bode \etal (2001); here $z$ is the redshift, 
$h$ is the Hubble parameter in unit of 100 (km/s)Mpc$^{-1}$, $\Omega_{\rm w}$, $m_{\rm w}$ and $g_{\rm w}$ are the density parameter, the mass and the number of spin states of WDM respectively.
Their distribution function reads then $(e^{{p_{\rm w}}/{T_{\rm w}}}+1)^{-1}$ ($p_{\rm w}$ and $T_{\rm w}$  being the momentum and temperature of WDM), until
gravitational clustering begins (Bode \etal 2001).

\subsection{Large Box simulations}

We 
run two kinds of simulations, the first in a set of fairly large
cosmological boxes with the same resolution across the whole box. We
have three different volumes 
with $L = 20,~ 40,~ 90 \Mpch$ aside, each of them containing $300^3$
dark matter particles. For each box we 
run two simulations, one with our SCDEW model, and a reference one
done in standard \LCDM for the same set of cosmological parameters (see
table \ref{tab:param} for a complete list).  We have used same values
for the Hubble constant ($h=0.685$), the spectral index ($n_s=0.968$)
and the power spectrum normalization ($\sigma_8 = 0.833$) in all
models.

\begin{table}
\begin{center}
  \caption{Cosmological parameters for the \LCDM and SCDEW models. Here 
$\Omega_{\Phi}$, $\Omega_{c}$, $\Omega_{\rm w}$ and  $\Omega_{b}$ are
the density parameters of DE,CDM, WDM and baryons. All models
  share the same values for the Hubble parameter ($h=0.685$), the spectral
  index ($n_s=0.968$) and the power spectrum normalization ($\sigma_8 = 0.833$).}
\label{tab:param}
\begin{tabular}{lccccc}
\hline
\hline
Label & $\Omega_{\Phi}$  & $\Omega_{c}$ & $\Omega_{\rm w}$  &  $\Omega_{b}$  & $m_{\rm w}$ [eV]  \\ 
\hline
SCDEW & 0.704 & 0.001 & 0.250 & 0.045 &  90 \\ 
\LCDM & 0.704 & 0.251 & -- & 0.045 &  -- \\ 
\hline
\end{tabular}
\end{center}
\end{table}

In all simulations, dark matter haloes are identified using a
spherical overdensity (SO) algorithm as described in Dutton \&
Macci\`o (2014). For our \LCDM cosmology the virial density at redshift
zero is $\Delta(0) \simeq 95.0$ based on the fitting function of
Mainini \etal (2003); we have used the same value for the SCDEW model
as well.

For each SO halo in our sample
we evaluated
the concentration parameter following the procedure outlined in
Macci\`o \etal (2008).  Briefly, we first determine the halo density
profile, by using the most bound particle as the location of the halo
center. We then compute the density ($\rho_i$) in 50 equally spaced
(in log) spherical shells. The minimum radius is the maximum between
1\% of the virial radius, $\Rvir$, or 3 times the softening length and the maximum radius
is $1.2\Rvir$. Errors on the density are estimated from the Poisson
noise due to the finite number of particles in each mass shell. We fit
these density profiles with the NFW expression (Navarro \etal 1997):
\begin{equation}
\rho_{\rm NFW}(r) = \frac{\delta_c \rho_{\rm cr}}{(r/r_{s})(1+r/r_{s})^2};
\label{eq:nfw}
\end{equation}
here $r_{s}$ is the scale radius of the halo, $\delta_c$ is
normalization parameter, and $\rho_{cr}$ is the critical density of
the Universe. Their values, and associated uncertainties, are obtained
via a $\chi^2$ minimization procedure using the Levenberg \& Marquart
method. We define the r.m.s. of the fit as
\begin{equation}
\rho_{\rm rms} = \sqrt{\frac{1}{N}\sum_i^N { (\ln \rho_i - \ln \rho_{\rm m})^2}}~,
\label{eq:rms}
\end{equation}
$\rho_{\rm m}$ being the fitted NFW density distribution.
Finally, following Macci\`o \etal (2007), we only selected relaxed
haloes
by requiring $\rho_{\rm rms} < 0.5$ and $x_{\rm off} < 0.07$. 
Here $\rhorms$ is the r.m.s. of the NFW fit to the density profile and
$\xoff$ is the offset between the most bound particle and the center
of mass, in units of the virial radius $\Rvir$.

\subsection{High resolution zoomed simulations}

We 
%have 
then selected a few haloes to be re-run at much higher resolution,
so allowing a more detailed study of the effect of our SCDEW models on
the inner structure of haloes (e.g. density profiles) and on their
satellite population.

One halo comes from the 90 \Mpch box and 
has a
mass similar to our own Milky-Way ($\approx 10^{12} \hMsun$). We will
refer to it as MW1. This halo was run with a mass resolution increase
of 4096, reaching a mass per particle of $5.41\times 10^5 \hMsun$ and
a softening of 0.4 \hkpc. For this halo we performed three different
runs, a \LCDM one, two SCDEW, either with thermal velocities or without
them.  In order to identify bound subhaloes we used the {\sc
  ahf}\footnote{The Amiga Halo finder ({\sc ahf}) can be freely
  downloaded from http://www.popia.ft.uam.es/AMIGA} halo finder
(Knollmann \& Knebe 2009).

Two further haloes (D1 and D2) have a much lower mass, similar to dwarf galaxies
($\approx 5 \times 10^{9} \hMsun$), and have been selected from the 20
\Mpch box. They were also zoomed in by a factor of 4096, obtaining in
this case a mass per particle of $5.94\times 10^3 \hMsun$ and a
softening of 0.1 \hkpc.

Unfortunately, no \LCDM counterpart of these haloes could be run.
As a matter of fact,
on such small scales there are a lot of modifications to the tidal
gravitation field when moving from \LCDM to SCDEW, which alter the
evolution pattern
of the same initial Lagrangian region. In order to have a \LCDM
analogous of our dwarf SCDEW galaxies, we then turned to the 20 \hMpc
box in \LCDM and selected there a halo with mass and environment
similar to the halo D2. We run it at high resolution, and will call it
D1L. All the parameters of high resolution halos are summarized in
table \ref{tab:sim}.

\begin{table}
\begin{center}
\caption{High resolution simulations properties.}
\label{tab:sim}
\begin{tabular}{llccc}
\hline
\hline
Halo & Model & $M_{\rm vir} $  & $N_{\rm vir}$ & Softening   \\ 
 &   & $[\hMsun]$  &   & [\hkpc]   \\ 
\hline
MW1 & SCDEW  & $8.51  \times 10^{11}$& 1,573,013 & 0.4 \\
MW1 & \LCDM  & $9.11  \times 10^{11}$  & 1,685,767 & 0.4 \\
D1 & SCDEW  & $8.92  \times 10^{9}$& 1,501,683 & 0.1 \\
D2 & SCDEW  & $1.31  \times 10^{10}$& 2,205,387 & 0.1 \\
D1L & \LCDM  & $7.51  \times 10^{9}$& 1,264,309 & 0.1 \\

\hline
\end{tabular}
\end{center}
\end{table}

%%%%%%%%%%%%%%%%%%%%%%%%%%%%%%%%%%%%%%%%%%%%%%%%%%%%%
\section{Results}\label{sec:Results}
%%%%%%%%%%%%%%%%%%%%%%%%%%%%%%%%%%%%%%%%%%%%%%%%%%%%%

We will start by presenting the analysis of large box simulations.
Figure \ref{fig:mf} shows the halo mass function in the two different
models (obtained by combining all three different boxes of sizes
20,~40,~90 \hMpc). As expected from the Power Spectrum behavior (see
figure \ref{fig:power}), the two mass functions agree at high masses
(large scales), while exhibiting some discrepancy as the halo mass
decreases. At intermediate masses the \LCDM model (blue line) produces
more haloes with respect to the SCDEW one, while at the lowest masses
probed by our simulations ($M\approx 10^{10} \hMsun$) there is an
excess of haloes in the SCDEW model. Differences are small anywhere,
never exceeding a factor of two.

\begin{figure}
\includegraphics[height = 0.5\textwidth]{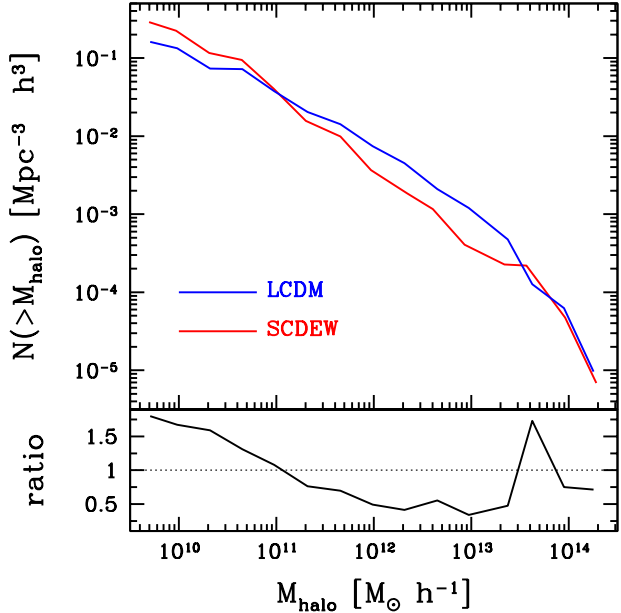}
\caption{Halo Mass function at redshift zero: \LCDM is shown in red, while
  SCDEW is shown in blue. The lower panel shows the ratio between the SCDEW and the \LCDM
mass functions.}
\label{fig:mf}
\end{figure}

The concentration--mass plot shown in figure \ref{fig:cm} exhibits
more significant differences between the two models. This is expected,
since the concentration parameter is more sensitive than the halo mass
function to changes in the cosmological background (Macci\`o \etal
2008). While SCDEW shows concentrations similar to \LCDM at the highest
and lowest masses we probe, SCDEW exhibits significantly lower
concentrations at intermediate masses: $5\times 10^{10}-5 \times
10^{13}$ \hMsun. These lower concentrations are due to lack of power on those
scales, in the SCDEW model; haloes then form at a lower redshift, so
determining lower concentrations (e.g. Wechsler \etal 2002).

\begin{figure}
\includegraphics[height = 0.5\textwidth]{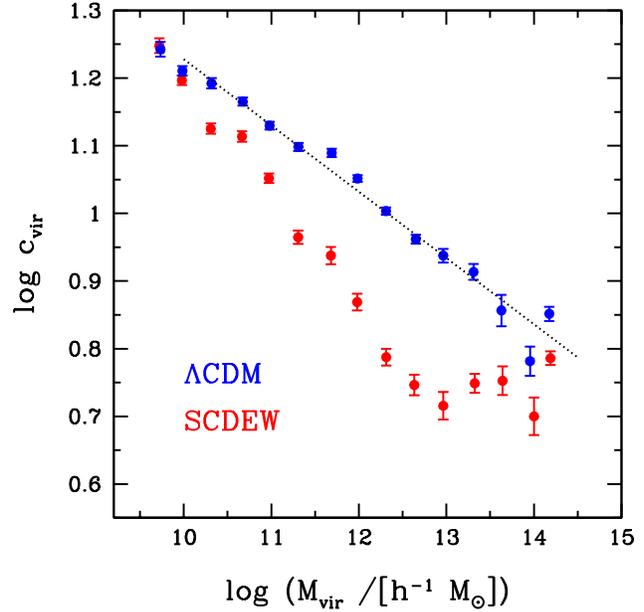}
\caption{The concentration mass relation at $z=0$. The blue points represent
  \LCDM, while the red ones show SCDEW. The black dotted line shows a simple
  linear fit (in log-log space) to the \LCDM results.}
\label{fig:cm}
\end{figure}

As already noted in the Introduction, observations of low surface
brightness galaxies and dwarf galaxies do suggest lower concentrations
with respect to the predictions of a standard \LCDM model (e.g. de Blok
et al. 2001; van den Bosch \& Swaters 2001; de Blok \& Bosma 2002;
Swaters \etal 2003; Dutton \etal 2005; Gentile \etal 2005; Simon \etal
2005) and hence they favor SCDEW.

\subsection{Structure of Milky Way-like haloes in SCDEW}

We will now study in more details the dark matter distribution and the
satellites population in our two Milky Way-like high resolution dark
matter haloes.

Figure \ref{fig:mw} shows a logarithmic density map of the three
versions of the MW1 halo: \LCDM (left), SCDEW without initial thermal
velocities (center) and SCDEW with thermal velocities (left). Even at
naked eye, we notice a clear difference among the abundances of
subhaloes in the three different cases.

\begin{figure*}
\includegraphics[height = 0.32\textwidth]{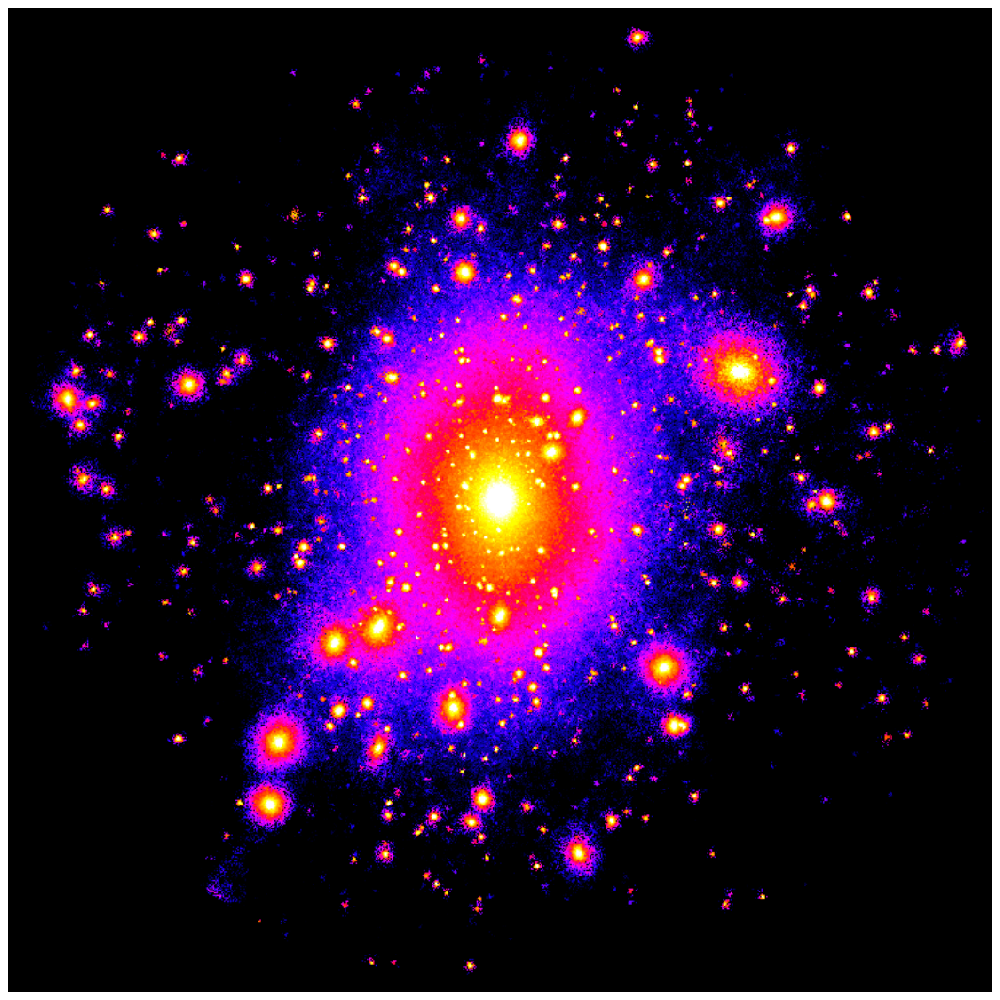}
\includegraphics[height = 0.32\textwidth]{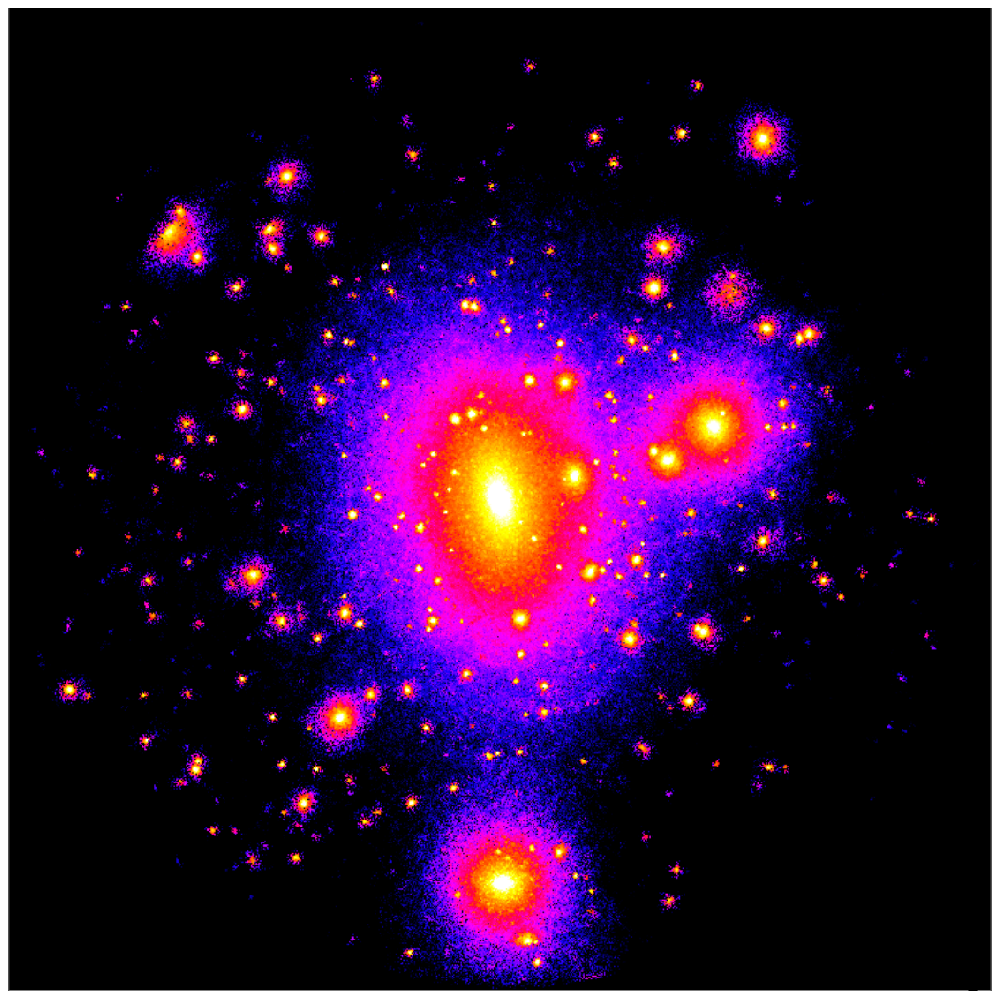}
\includegraphics[height = 0.32\textwidth]{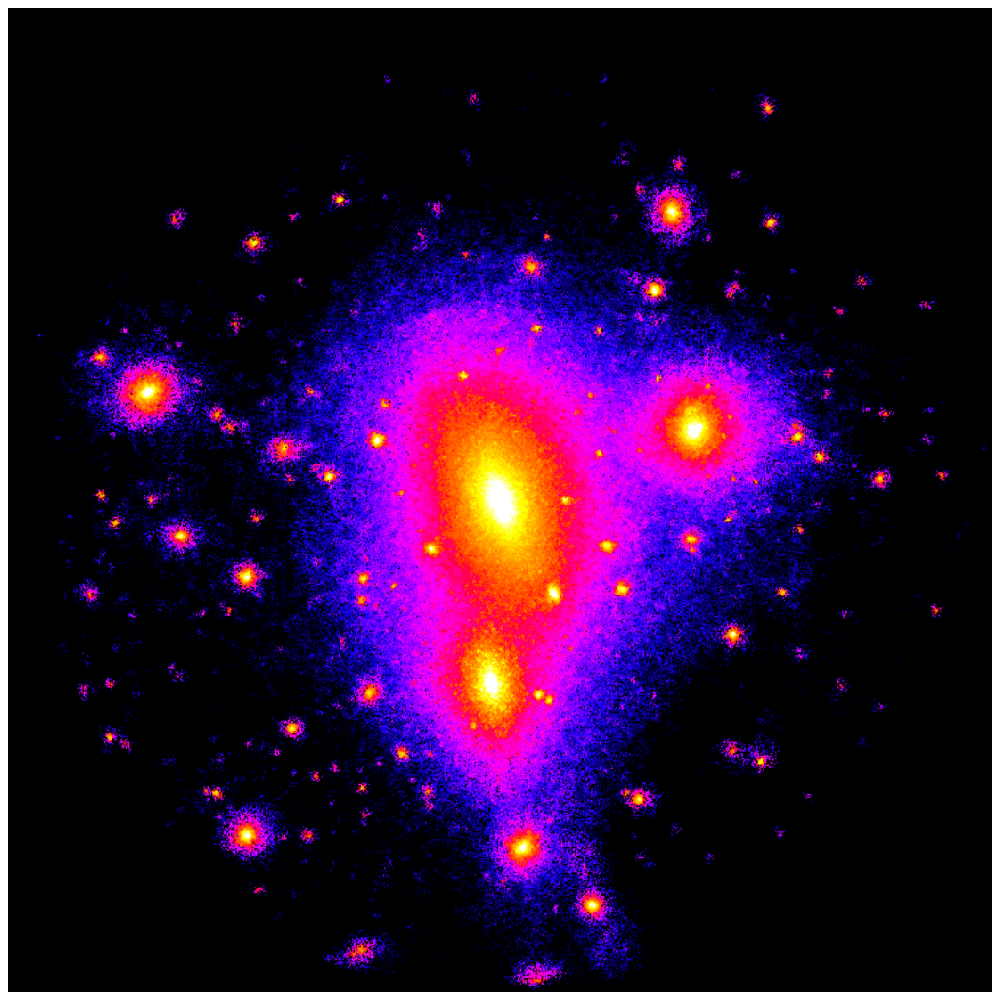}
\caption{Logarithmic density map of the MW1 halo in the different models, each panel
  is 400 \hkpc across, encompassing the halo virial radius.
  Left \LCDM, center SCDEW without
  thermal velocities, right SCDEW with thermal velocities.}
\label{fig:mw}
\end{figure*}

These differences are quantified in figure \ref{fig:submf1} where we
show the subhalo mass function for the above three simulations.  The
lower number of satellites in SCDEW with respect to \LCDM is due to the
combination of two effects: (i) There is  less power on the MW scale;
this delays the formation of the halo and lowers the environment
density at the formation time; in turn, this reduces the subhaloes
fraction (e.g. Maulbetsch \etal 2007). (ii) Furthermore, thermal
velocities are high (WDM particles are 90 eV~!) and this also
suppresses the formation of low mass structures. 

A comparison of the three lines in figure \ref{fig:submf1} clearly
shows that on the scale of the Milky Way, the former effect is more
important, while thermal velocities only affect the tail of the
subhalo mass function at masses $M<10^8 \hMsun$.

The subhalo mass function in the SCDEW model is quite different from
a pure Warm Dark Matter model, with the same WDM particle mass
$m_{\nu} = 90 eV$. In fact, in the latter case, the formation of {\it
  all} substructures
is suppressed, since the free streaming mass for such a model is of
the order of few $10^{14} \hMsun$ (Zentner \& Bullock 2003).
On the contrary, in the SCDEW model, thanks to the coupling between
CDM and DE, there are still enough substructures to host the formation
of dwarf galaxies and, hence, not to violate constraints on the
luminosity function of MW satellites (e.g. Macci\`o \& Fontanot 2010).

\begin{figure}
  \includegraphics[height = 0.5\textwidth]{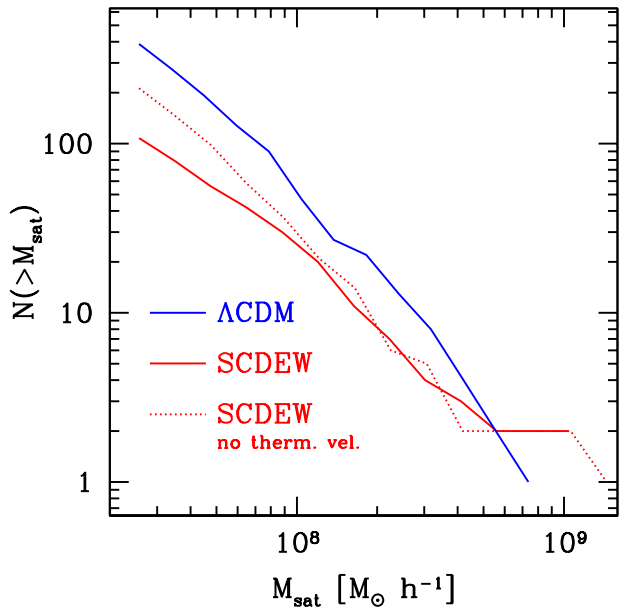}
  \caption{Subhalo integral mass function for the MW1 halo. \LCDM is shown in blue, while
    the solid and dotted lines represent SCDEW with and without thermal velocities respectively.}
\label{fig:submf1}
\end{figure}

Figure \ref{fig:MWprof} shows the radial density profile for our three
realizations of the MW1 halo. The \LCDM profile (blue) is slightly
higher in normalization w.r.t SCDEW, while sharing a similar slope;
the virial concentration parameters are 10.4 for \LCDM and 7.1 for
SCDEW; the difference is mainly due to the slightly larger virial
radius in the \LCDM run. The difference between the SCDEW runs with and
without velocities is almost negligible. This is not surprising
given the halo mass and our choice of WDM mass (90~eV), we
expect an effect on scales of few hundreds pc (e.g. Macci\`o \etal
2012), well below the resolution of our simulation. As we will see
later, the situation is quite different for dwarf galaxies.

\begin{figure}
\includegraphics[height = 0.5\textwidth]{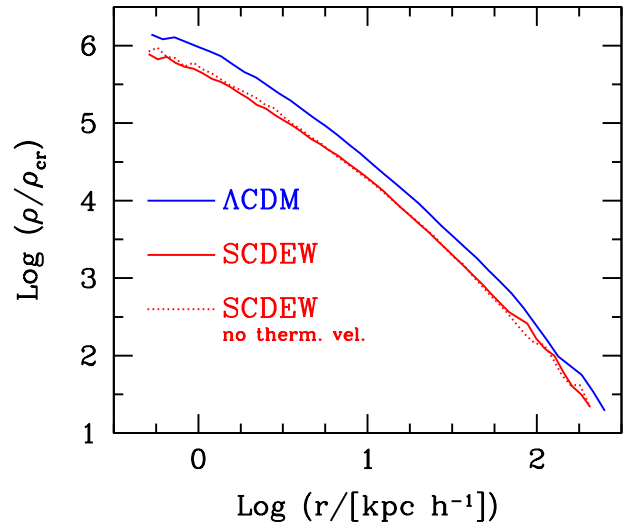}
\caption{Density profile for the MW1 halo. \LCDM is shown in blue, while
    the solid and dotted lines represent SCDEW with and without thermal velocities respectively.}
\label{fig:MWprof}
\end{figure}

\subsection{Structure of dwarf haloes in SCDEW}

In order to investigate the effects of SCDEW on small spatial and mass
scales we performed zoomed simulations of isolated dwarf haloes with
masses of $M_{\rm D1} = 8.92 \times 10^{9} \hMsun$ and $M_{\rm D2} =
1.31 \times 10^{10} \hMsun$.
As already mentioned, due to the large
differences in the tidal field between SCDEW and \LCDM for such small
scales, neither D1 nor D2 could be run in $\Lambda$CDM. We rather selected a
different halo with a similar mass $M_{\rm D1L} = 7.51 \times 10^9
\hMsun$ from the \LCDM simulation, and run it at the same resolution;
we dubbed it D1L.

Figure \ref{fig:dwarf} shows the density map of the D1 halo with and
without thermal velocities. At variance from what happened on the MW
scale, in the SCDEW run with thermal velocities all subhaloes have
disappeared, and only the central halo has survived. This sets an
effective free streaming mass $\sim 10^{7} \hMsun$ for our model,
equivalent to a warm dark matter (thermal) candidate of a few keV.

\begin{figure*}
\includegraphics[height = 0.47\textwidth]{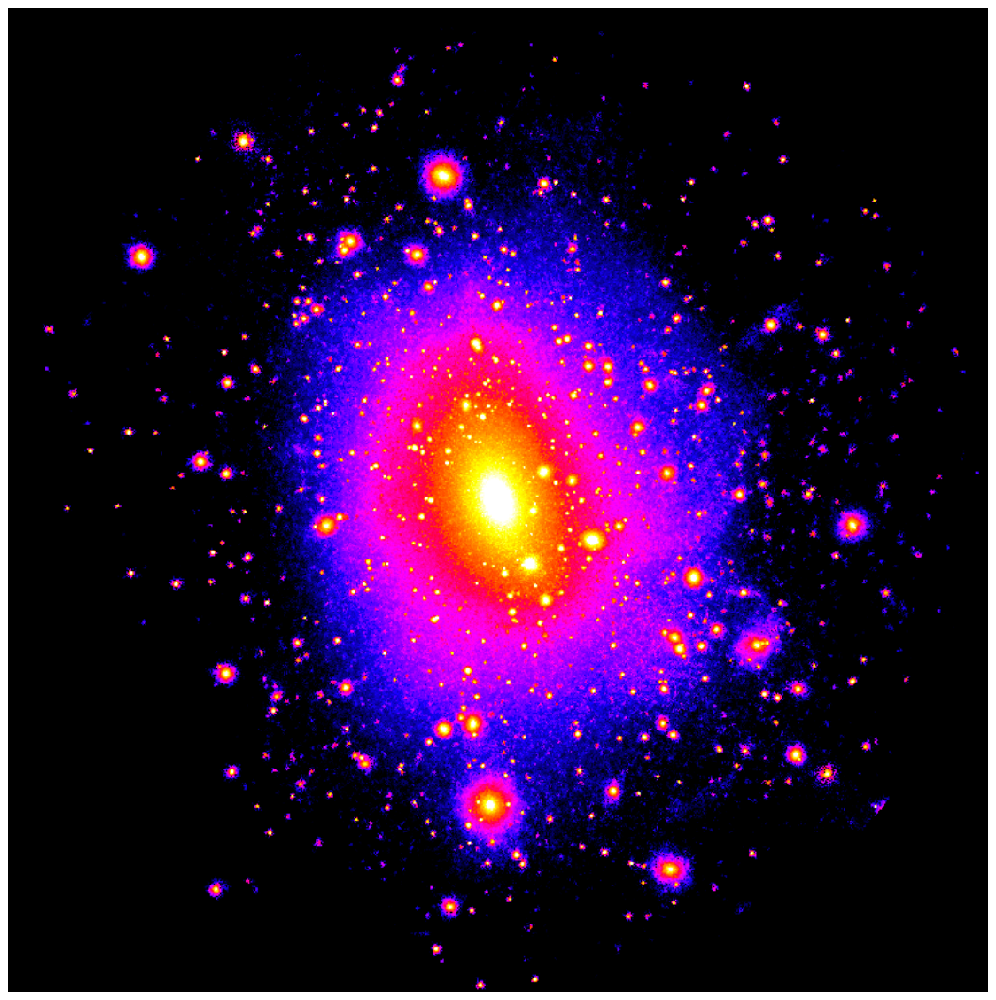}
\includegraphics[height = 0.47\textwidth]{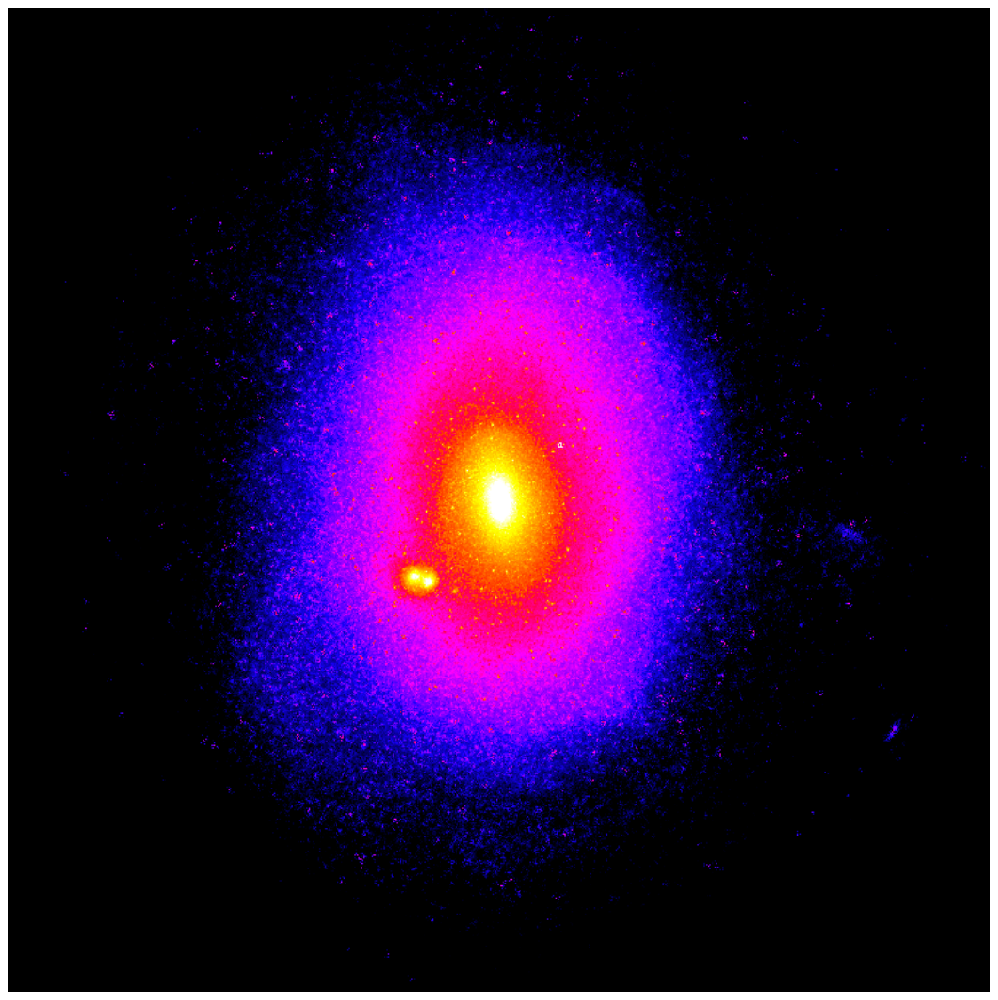}
\caption{Density map, in logarithmic scale for the halo D1,
each panel is 80 \hkpc across, encompassing the halo virial radius.
Left SCDEW with out thermal velocities, right SCDEW with thermal velocities}
\label{fig:dwarf}
\end{figure*}
    
The most interesting effect of the presence of a warm component can be
seen on the density profile of our dwarf galaxies. Figure
\ref{fig:profD1} shows the density profile of the D1 (red) and the D1L
(blue) haloes. The D1L profile exhibits the usual NFW--like shape with
a cuspy central slope; the SCDEW profile {\it without} thermal
velocities shows a NFW like behavior as well, with a cusp in the
center, even if at a lower normalization.

\begin{figure}
\includegraphics[height = 0.5\textwidth]{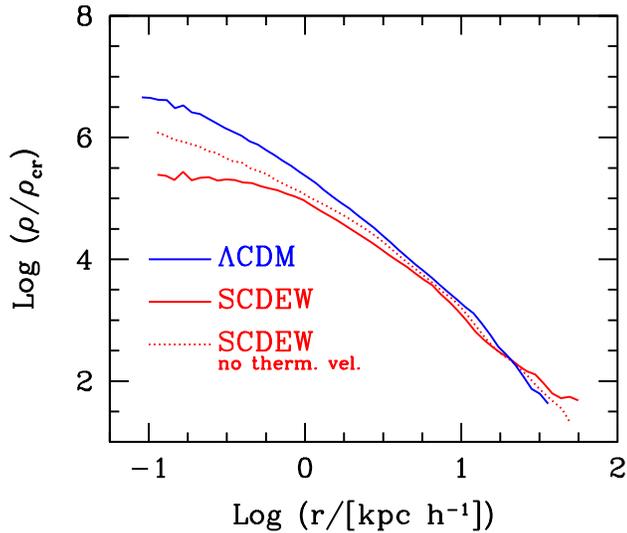}
\caption{Density profile for the dwarf galaxy halo D1L (blue) and the D1 with velocities (solid red)
  and without them (dotted red).}
\label{fig:profD1}
\end{figure}

The SCDEW halo {\it with} thermal velocities, instead, shows a very
clear {\it cored} profile. The size of the core is of the order of
several hundreds of pc, in agreement with theoretical expectations
based on the conservation of phase space density (Macci\`o \etal 2012,
Shao \etal 2013).
Halo D2 exhibits a similar behavior, as shown in figure
\ref{fig:profD2}. The central slope of the profile is $\alpha =
-0.25$, substantially shallower than the NFW prediction ($\alpha =
-1$), even though the core is less pronounced. In this plot the blue
line represents the theoretical expectation for an NFW profile in
\LCDM, assuming an average concentration for the halo D2 mass.

\begin{figure}
\includegraphics[height = 0.5\textwidth]{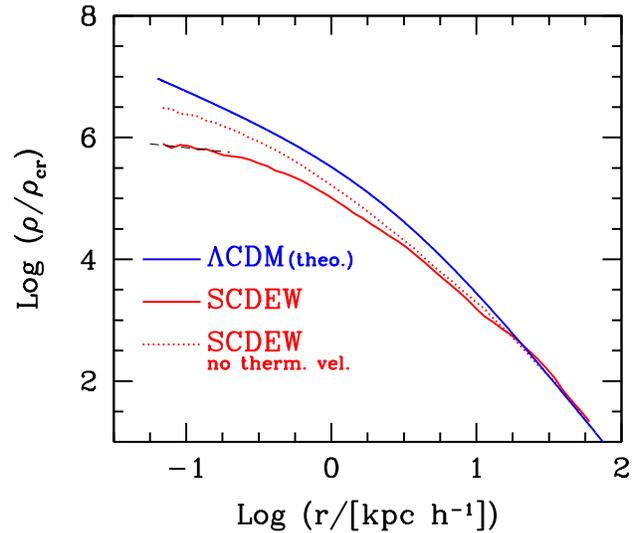}
\caption{Density profile for the dwarf galaxy halo D2. SCDEW is shown
  as solid red, while SCDEW without velocities is shown by the dotted
  red line.  The blue line is the \LCDM theoretical expectation, for an
  NFW profile at this halo mass with an average concentration from
  figure \ref{fig:cm}.}
\label{fig:profD2}
\end{figure}

The flattening of the DM profile and the creation of a core on such
small mass scales is one of the successes of the SCDEW model. Attempts
of measuring the density profile of DM in the milky Way satellites
suggests the presence of central cores (Walker \& Pe{\~n}arrubia 2011,
Amorisco \& Evans 2012, Amorisco \etal 2013) even though there is
still some debate on whether cores are really needed to reproduce the
kinematics of these satellites (Strigari \etal 2014). If cores are
indeed present in such small objects, it is quite unlikely that their
creation can be ascribed to baryonic effects (Garrison Kilmer \etal
2013), since the capability of baryons to modify the DM density
profile is linked to the stellar/DM ratio, which is extremely low in
these objects (Di Cintio \etal 14a).

Any cosmological model able to create cores at this mass scales will
definitely ease the tension with the observations and hence should be
preferred to a simple \LCDM model.

%%%%%%%%%%%%%%%%%%%%%%%%%%%%%%%%%%%%%%%%%%%%%%%%%%%%%
\section{Conclusions}\label{sec:conclusion}
%%%%%%%%%%%%%%%%%%%%%%%%%%%%%%%%%%%%%%%%%%%%%%%%%%%%%

The \LCDM model (based on DE with state parameter $w \equiv -1$ and
cold dark matter) is very successful in explaining the evolution of
our Universe and its redshift zero structure (e.g. Tegmark \etal
2007).

Despite its success, this model faces problem on the theoretical side,
due to its inability to explain the very low value of DE density
which, in turn, allows dark matter and dark energy densities to be
quite similar at the present day. Furthermore, such low value allowed
DE to become relevant late enough, so to allow primeval fluctuations
to turn into non--linear structures up to quite large scales.

The \LCDM scenario is also challenged by the possible presence of cores
in the observed dark matter distribution of Milky Way satellites
(e.g. Walker \& Pe{\~n}arrubia) since on such scales is very hard to
invoke baryonic effects to reconcile observations and collisionless
simulation predictions (Di Cintio \etal 2014a).

It is then worth to explore alternatives to the simple \LCDM model,
possibly trying to find a model that can deal, at the same time, with
both theoretical and observational issues.

In this paper we presented a detailed non linear analysis of strongly
coupled cold+warm cosmologies (SCDEW), while the linear theory and the
basic features of these models have been presented in the companion
paper (Bonometto, Mainini \& Macci\`o 2015, Paper I).

Thanks to the coupling between Dark Energy and Cold Dark Matter, these
models are able to ease the fine tuning and, possibly, the coincidence
problem depending on the model parameters, problems that plage a simple \LCDM scenario.
A very important feature of
SCDEW is that the (uncoupled) warm dark matter component can be chosen
with very low particle mass, as low as $\sim 100$~eV, since the
strongly coupled CDM is able to regenerate fluctuations in the WDM
component after its derelativisation (see Paper I).  As a consequence,
the power spectrum of the SCDEW cosmology is very similar to the one
of standard $\Lambda$CDM, and hence in very good agreement with observations.

We have run and analyzed several Nbody simulations performed in the
SCDEW scenario and compared them with analogous simulations in the
\LCDM model. On large scales, SCDEW is quite similar to \LCDM in halo
distribution and number density; this implies that SCDEW shares the
same success of \LCDM on these scales. On the other hand, the
distribution of mass in collapsed objects is less concentrated in
SCDEW than in \LCDM easing possible tensions with observations
(e.g. Salucci \etal 2012).

By means of high resolution simulations, we then studied in detail the
properties of haloes on the scales of our own Galaxy (the Milky Way)
and dwarf galaxies.  On MW scales, the SCDEW halo presents less
substructures (30-40\%) than its \LCDM counterpart, possibly helping
any ``baryonic'' solution (e.g. Benson \etal 2002, Macci\`o \etal
2010) of the so called missing satellites problem (e.g. Klypin \etal
2001). The total density profile is still well represented by an NFW
fit, even though with a lower concentration parameter.

Drastic improvements on \LCDM are however found on the scale of dwarf galaxies
Thanks to the large initial thermal velocities of the warm particles
(due to the very low WDM particle mass of $90$~eV), the density
profile in SCDEW haloes has a clear central core. The size of the core
is of the order of several hundreds of parsecs, in good agreement with
observations of the MW satellites like Sculptor and Fornax
(e.g. Amorisco \etal 2013).

SCDEW models are still in an infant state and more simulations and
theoretical work is needed to fully test them at the same level as \LCDM.
For example we have negletted the possible formation of (very small) collapsed objects in
  CDM before the fading of the coupling. While these objects are expected to not sensible
  perturb the main gravitation potential, they could still leave some signature
  in the observational data (e.g. secondary CMB spectrum) and ought to be included
  in future numerical studies.

Nevertheless our theoretical (Paper I) and Numerical (this
paper) works clearly show that strongly coupled cold-warm cosmologies
are able to preserve the success of \LCDM on large scales and to {\it
  strongly} improve the agreement with data on very small scales,
simultaneously easing the theoretical short-comes of DE in $\Lambda$CDM,
% a cosmological constants, 
and thus appearing as a valid alternative to a simple \LCDM scenario.

%%%%%%%%%%%%%%%%%%%%%%%%%%%%%%%%%%%%%%%%%%%%%%%%%%%
\section*{Acknowledgments}

AVM and CP acknowledge support from 
the  Sonderforschungsbereich SFB 881 ``The Milky Way System'' 
(subproject A2) of the German Research
Foundation (DFG).  
We acknowledge the computational support from the {\sc theo} and {\sc hydra} 
clusters of the
Max-Planck-Institut f\"{u}r Astronomie at the Rechenzentrum in
Garching.
SAB thanks the C.I.F.S. (Consorzio Interuniversitario per la Fisica
Spaziale) for its financial support.

%%%%%%%%%%%%%%%%%%%%%%%%%%%%%%%%%%%%%%%%%%%%%%%%%%%

%%%%%%%%%%%%%%%%%%%%%%%%%%%%%%%%%%%%%%%%%%%%%%%%%%%%%%%%%%%%%%%%%%%%%%
%%  REFERENCES
%%%%%%%%%%%%%%%%%%%%%%%%%%%%%%%%%%%%%%%%%%%%%%%%%%%%%%%%%%%%%%%%%%%%%% 

%\bibliographystyle{mn2e}
%%\bibliography{archive}

%\bsp

%\label{lastpage}

\end{document}